\documentclass[prl,twocolumn,showpacs,superscriptaddress,floatfix]{revtex4}
\usepackage{graphicx,amsfonts,amssymb,amsmath,times}
\usepackage[colorlinks,bookmarks=false,citecolor=blue,linkcolor=red,urlcolor=blue]{hyperref}
\usepackage{color}

\newlength{\ldag}
\begin{document}

\title{Family of exactly solvable models with an ultimative quantum paramagnetic ground state}

\author{Kai Phillip Schmidt}
\email{schmidt@fkt.physik.uni-dortmund.de}
\affiliation{Lehrstuhl f\"ur Theoretische Physik I, Otto-Hahn-Stra\ss e 4, TU Dortmund, D-44221 Dortmund, Germany}

\author{Mukul Laad}
\email{laad@fkt.physik.tu-dortmund.de}
\affiliation{Lehrstuhl f\"ur Theoretische Physik I, Otto-Hahn-Stra\ss e 4, TU Dortmund, D-44221 Dortmund, Germany}

%------------------------------------------------------------------------------

\begin{abstract}
  We present a family of two-dimensional frustrated quantum magnets solely based on pure nearest-neighbor Heisenberg interactions which can be solved quasi-exactly. All lattices are constructed in terms of frustrated quantum cages containing a chiral degree of freedom protected by frustration.  The ground states of these models are dubbed ultimate quantum paramagnets and exhibit an extensive entropy at zero temperature. We discuss the unusual and extensively degenerate excitations in such phases. Implications for thermodynamic properties as well as for decoherence free quantum computation are discussed.          
\end{abstract}


\maketitle
% 
%
%%%%%%%%%%%%%%
%\emph{Introduction ---}
%%%%%%%%%%%%%%
%
%
The search for exotic phases of strongly correlated quantum matter possessing unusual physical properties is a fascinating 
and active research area. Exactly solvable models played an essential role toward this end.  Important examples like 
the Heisenberg chain or the frustrated Majumdar-Ghosh point with free or gapped spinon excitations exist in one dimension \cite{Cloizeaux62,Faddev81,Majumdar69}. 
In two dimensions, the frustrated Shastry-Sutherland model constitutes the paradigm example for a valence bond 
solid \cite{Shastry81}. Its experimental realization in SrCu$_2$(BO$_3$)$_2$ has spurred intense activity. Finally, the 
exactly solvable toric code \cite{Kitaev03,Wen03} and Kitaev's honeycomb model \cite{Kitaev06} have become 
standard models rigorously exhibiting topologically ordered phases. These have received enormous attention in the context of topological 
quantum computation.

  As already indicated above, one driving knob to tune quantum fluctuations 
%and therefore to surpress conventional classical order 
is geometrical frustration. Indeed, the most promising candidates for experimental realizations of a quantum spin liquid (SL) 
are strongly frustrated systems like the so-called Herbertsmithite \cite{Helton07, Mendels07} or the undoped parent compound 
$\kappa$-(BEDT-TTF)$_{2}$Cu$_{2}$(CN)$_{3}$ of the organic superonductors \cite{Shimizu03}. 
In Herbertsmithite, the geometrical frustration in the Heisenberg model on the kagome lattice is proposed to stabilize a SL state \cite{Ran07}, while 
multi-spin interactions are relevant for the organic compound \cite{Motrunich05}.   

Theoretically, different quantum SL possessing very different physical properties have been proposed. Gapped SL have been exactly demonstrated in the quantum dimer model on non-bipartite lattices \cite{Moessner01, Misguich02} and in the toric code \cite{Kitaev03,Wen03}. Further interesting examples are the U(1) critical SL discussed in the framework of the kagome model \cite{Ran07} and Bose-metal phases discussed for triangular topologies with multi-spin interactions \cite{Sheng08}. Finally, there are intriguing proposals for so-called ultimate co-operative paramagnets in context of the transverse field Ising model on the kagome lattice \cite{Moessner00}. In all these cases, no rigorous examples of such exotic phases in two dimensions are known. Here, we present a family of quasi-exactly solvable microscopic models having quantum disordered, extensively degenerate ground states and ultra short-ranged spin correlations. We dub these distinct class of SL ultimate quantum paramagnets (UQP).  

Our models might well be realized in experiment and, additionally, could be relevant for quantum information (QI). First, the models consist solely of pure nearest-neighbor spin-1/2 Heisenberg antiferromagnets (HAF), i.e.~coupled sites $i$ and $j$ interact via $\vec{S}_{i} \cdot \vec{S}_{j}$. This kind of exchange is realized in many physical systems. In the QI context, we are especially motivated by coupled quantum-dot systems \cite{Loss98,Kane98,Petta05}, possible patterning studies of quantum cellular automata (QCA)~\cite{Toth01}, trapped ions~\cite{Porras04,Friedenauer08}, and by using triangular magnetic clusters like Cu$_{3}$~\cite{Trif08,Georgeot09}.  Suitably designed Josephson junction arrays (JJA) are also of interest here, since the JJA can be mapped onto $S=1/2$ Heisenberg-like models~\cite{Bradley84}.  We draw attention to the fact that the core element in QCA computation is a bistable ``cell'' capable of interacting with its neighbors, which is precisely the strategy used here. Indeed, protected chiral quantum bits as also engineered in our case in a scalable fashion have been the focus of recent works in QI \cite{Trif08,Georgeot09}.    

\begin{figure}
\begin{center}
\includegraphics*[width=1.0\columnwidth]{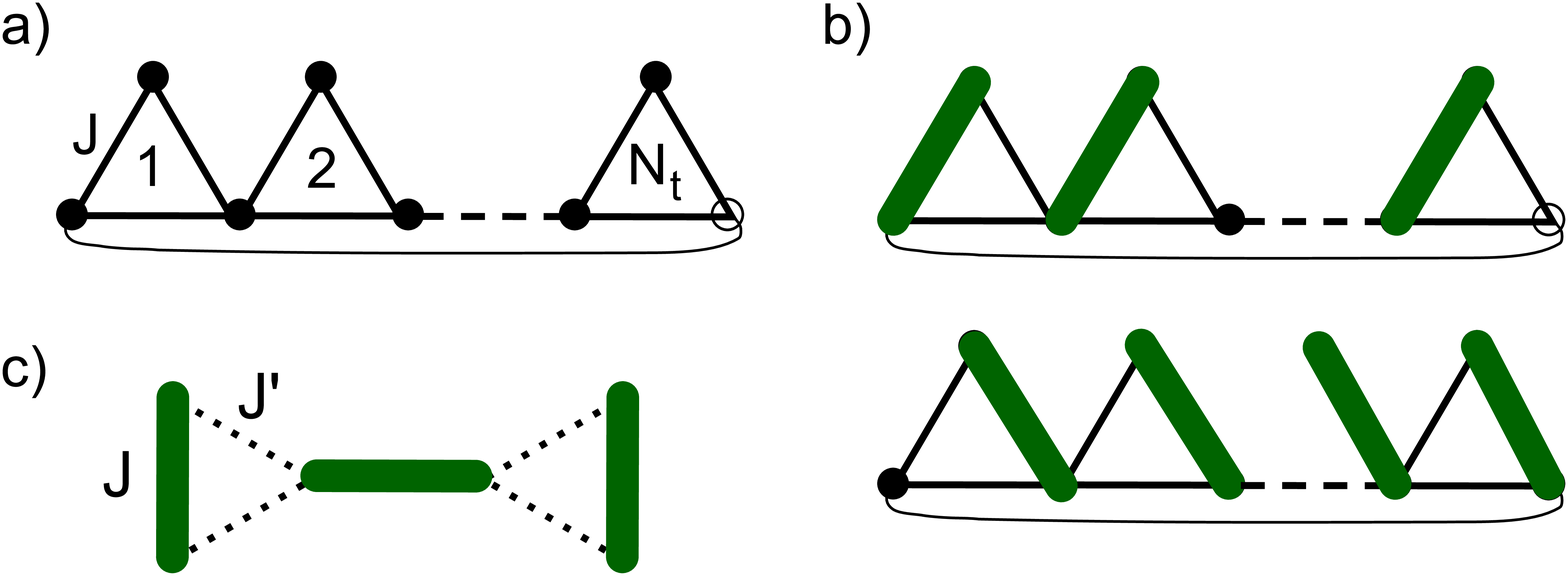}
\end{center}
\caption{(Color online) (a) Heisenberg model with exchange $J$ on a periodically coupled chain of $N_{\rm t}$ triangles. (b) Illustration of the two degenerate ground states. Thick (green) bonds represent singlet states. 
%The two ground states are singlet coverings oriented either to the right or to the left. 
(c) The shortest segment of a Shastry-Sutherland chain consists of two vertical and one horizontal dimer. Note that the two dimers on the edges are part of the quantum cage.}  
\label{fig1}
\end{figure}

% 
%
%%%%%%%%%%%%%%
%\emph{Construction of models ---}
%%%%%%%%%%%%%%
%
%
Let us first introduce the construction recipe for the lattices we study in this work. This is crucial since all interesting physical properties emerge 
from the special frustrated topology \cite{Schulenburg02}. The elementary building block is a finite chain of $N_{\rm t}$ coupled triangles with periodic boundary 
conditions, as illustrated in Fig.~\ref{fig1}a. Clearly, the ground state of this model is two-fold degenerate for $N_{\rm t}>2$. These states are 
singlet coverings which are either oriented to the right or the left (see Fig.~\ref{fig1}b). Thus, this two-level system represents a chiral pseudo-spin.
%Figure Illustration Triangular Chain
%%%%%%%%%%%%%%%%%%%%%%%%%%%%%%%%%%%%%%%%%%%%%%
%%%%%%%%%%%%%%%%%%%%%%%%%%%%%%%%%%%%%%%%%%%%%%

\begin{figure}
\begin{center}
\includegraphics*[width=0.8\columnwidth]{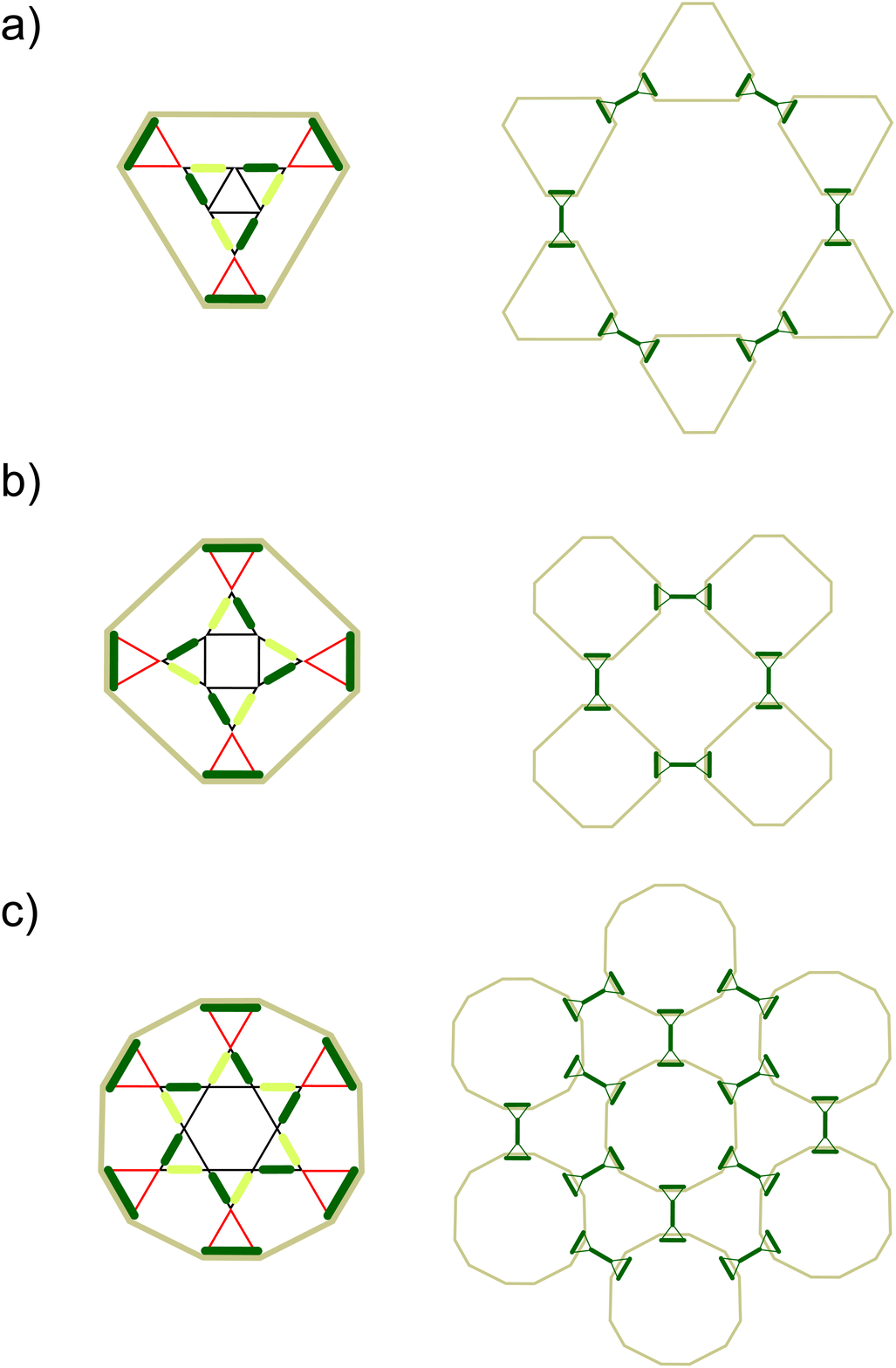}
\end{center}
\caption{(Color online) Three representative lattice models with (a) $N_{\rm t}=3$, (b) $N_{\rm t}=4$, and (c) $N_{\rm t}=6$. The left sides illustrates the chiral quantum cage of each lattice which contains a triangle chain shielded by surrounding dimers.
% The Heisenberg coupling between dimers is fully frustrated. 
The right sides represent 2d lattices made by adding dimers which couple cages. The dimer connecting two cages forms the shortest segment of a Shastry-Sutherland chain (illustrated in black).} 
\label{fig2}
\end{figure}

 Next, we put this chain segment on a ring as shown in Fig.~\ref{fig2} for $N_{\rm t}\in\{3,4,6\}$. Now we add $N_{\rm t}$ dimers 
on the boundary to form cage-like structures. Since the boundary dimers are coupled to the inner triangles in a Shastry-Sutherland fashion, singlets 
are formed on the outer dimers in the ground state of the full cage and the two-fold chiral degeneracy remains protected. Hence, we dub this unit a 
{\it chiral quantum cage}. 

The final step is to couple such cages. To this end, we introduce one additional dimer between two cages such that the inter-cage coupling involves the 
shortest segment of a one-dimensional Shastry-Sutherland chain (see Fig.~\ref{fig1}c) as illustrated in the right side of Fig.~\ref{fig2}.  The boundary dimers are coupled to the connector with strength $J'=xJ$. In this way, one can build any two-dimensional lattice of coupled cages. We are only aware of two works dealing with Hubbard \cite{Batista03} and Heisenberg \cite{Yang10} models on decorated lattices in the context of unconventional order, but gapped spin liquid ground states have never hitherto been considered in this context.

% 
%
%%%%%%%%%%%%%%
% Ground state
%%%%%%%%%%%%%%
%
%
  We want to study the ground states and elementary excitations as a function of $x=J'/J$. Our focus will be the regime $x$ $<$ $x_{\rm c}$ where, remarkably, the ground states of these models can be found {\it exactly}.  To see this, notice that the \emph{full} lattice can be covered by singlet configurations on each cage.  The inter-cage coupling involves the basic Shastry-Sutherland unit, i.e. two vertical singlets from neighboring cages are linked by one horizontal singlet.  
%For $x<x_{\rm c}$, this unit exactly frustrates an inter-cage coupling.  
Since, as a result, the chiral degree of freedom on neighboring cages are effectively decoupled, the exact ground states for $x<x_{\rm c}$ are a direct product of the dimer coverings on {\it each} of the $N_{\rm c}$ cages. In contrast to the dimer solid in the Shastry-Sutherland model \cite{Shastry81}, the crucial point here is that this number is \emph{extensively} large. The ground state degeneracy scales like $2^{N_{\rm c}}$, leading to a finite entropy $ln(2)$ per cage at $T=0$ in the thermodynamic limit. As $x$ increases, a phase transition at $x$ $=$ $x_{\rm c}$ will certainly occur (see below). 
%Figure Illustration Lattices
%%%%%%%%%%%%%%%%%%%%%%%%%%%%%%%%%%%%%%%%%%%%%%
%%%%%%%%%%%%%%%%%%%%%%%%%%%%%%%%%%%%%%%%%%%%%%

% 
%
%%%%%%%%%%%%%%
% Excitations
%%%%%%%%%%%%%%
%
%
But let us first discuss the elementary excitations in the UQP ground state for $x<x_{\rm c}$. They can also be determined quasi-exactly. 
% and have very intriguing character. 
 One finds three sorts of massive particles with total spin one above the infinitely degenerate ground state: 
i) excitations on the connector, ii) excitations on the boundary dimers, and iii) excitations inside the cage (stars). 

Excitations on the connector are localized triplets. This is a direct consequence of the peculiar Shastry-Sutherland 
geometry in one dimension which surpresses any fluctuations to neighboring dimers. The connecting dimer can be only in a singlet or a triplet state, both eigenstates of the full problem. The total Hilbert space separates into different blocks belonging to a different set of 
singlet-triplet configurations. A single triplet costs the exact energy $J$ independent of $x$. Clearly, the true low-energy excitations of the model belong to the sector with singlets on all connecting dimers.
% (see next paragraph). 

\begin{figure}
\begin{center}
\includegraphics*[width=0.9\columnwidth]{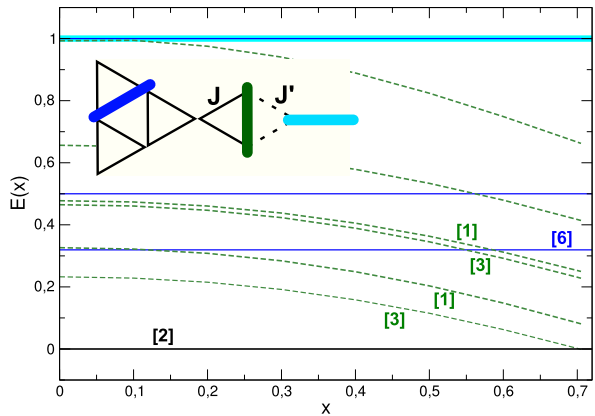}
\end{center}
\caption{(Color online) Localized excitations on a chiral quantum cage with $N_{\rm t}=3$ as a function of $x$. Thick grey line at energy $J$ corresponds to the exact triplet on the connecting dimer. Blue horizontal lines represent excitations inside the star. Dashed lines involve localized triplons on the outer shell which are the only single modes depending on $x$. The lowest of these modes is the energy of a single triplon. Other dashed lines corresond to a triplon plus an excitation inside the star. Number in square brackets gives the degeneracy on the cluster shown in the inset which has been diagonalized.} 
\label{fig3}
\end{figure}

The second sort of excitations are on dimers on the boundary of a cage. These excitations also stay local.  But triplets are not exact eigenstates, since they can virtually create and annihilate triplets on the inner star as well as on the attached connector, thereby reducing its energy. These excitations are 
local triplons \cite{Schmidt03} and constitute the excitations with lowest energy. This mode is displayed in Fig.~\ref{fig3} for $N_{\rm t}=3$. One finds a gap 0.23J at $x=0$. The gap is lowered as $x$ increases and the mode crosses the low-energy chiral manifold at $x\approx 0.7$. This signals a phase transition as long as no other levels cross before. Below, we argue this to be the case here. 

Finally, one has the excitations inside the star of a cage. 
The lowest modes correspond to very weakly dispersive, massive triplons. Let us note that these excitations do not depend on $x$ and can therefore be identified as the horizontal gaps in Fig.~\ref{fig3} for $N_{\rm t}=3$. Studies on the infinite chain $N_{\rm t}=\infty$ found completely local excitations 
with an energy gap 0.22J \cite{Kubo93}. For finite $N_{\rm t}$ the energy gap is slightly larger. One finds 0.32J for $N_{\rm t}=3$. The weakly dispersing character is nicely understood from the dimerized limit of the inner star of a cage by introducing a modulation $J\pm\delta$ on the spikes of the triangles (see Fig.~\ref{fig1}). In the limit $\delta=1$ the ground state is the product state of singlets and excitations are local triplets. Turning on a finite $\delta$, a first hopping processes of the 
triplet occurs in perturbation order $\delta^{2(N_{\rm t}-1)}$. One first excites all triplets of the star one after the other ($N_{\rm t}-1$ operations) and subsequently annihilates $N_{\rm t}-1$ triplets such that the triplet has effectively hopped. Thus, one directly understands that the excitations become increasingly local with increasing $N_{\rm t}$.          

A fourth exactly known class of excitations is the combined creation of a triplon on the outer shell plus excitations of the star. These modes certainly depend on $x$ due to the presence of the local triplon. Interestingly, a strong attraction is found for certain modes which leads to the presence of levels at rather low energy (see Fig.~\ref{fig3}) \footnote{The energies of more triplons on the same outer shell plus an excitation inside the star are also known exactly.}.   

In sum, we have seen that ground states and all relevant low-energy excitations for $x$ $<$ $x_{\rm c}$ can be found quasi-exactly. Excitations on the outer shell of the cages have the lowest energy. We stress that these excitations are localized on the quantum cage. This implies an extensive degeneracy for these states as well. Since ground state correlations are strictly zero for distances larger than the extension of a quantum cage, we have found a family of exactly solvable models with UQP ground states. 

%Figure Excitations Nt=3
%%%%%%%%%%%%%%%%%%%%%%%%%%%%%%%%%%%%%%%%%%%%%%
%%%%%%%%%%%%%%%%%%%%%%%%%%%%%%%%%%%%%%%%%%%%%%

\begin{figure}
\begin{center}
\includegraphics*[width=0.9\columnwidth]{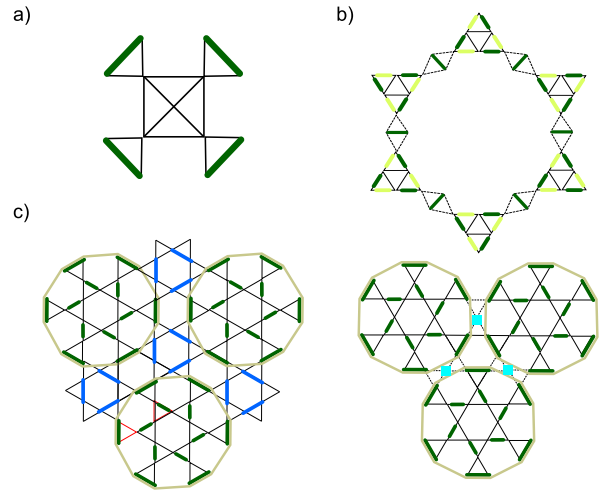}
\end{center}
\caption{(Color online) a) The fully-frustrated plaquette is the smallest inner star. b) A direct coupling of stars ($N_{\rm t}=3$ shown) leads to a different family of models. c) Illustration of the two possible close packings of $N_{\rm t}=6$ quantum cages on the Kagome lattice. Left figure represents the proposed valence bond solid ground state with the 36-site unit cell. Right figure represent a diluted Kagome lattice where sites marked by grey squares are removed.}  
\label{fig4}
\end{figure}

% 
%
%%%%%%%%%%%%%%
% Other set ups - plaquette- direct star connector
%%%%%%%%%%%%%%
%
%
This can actually be extended in two respects. First, one can change 
%the structure of 
the stars to different geometries also possessing a two-fold degenerate ground state, e.g. a fully-frustrated plaquette
%, illustrated in Fig.~\ref{fig4}a
with the minimal number of four spins. If one couples the fully frustrated plaquettes in a completely analogous fashion as before, one again finds exactly an UQP ground state for $x<x_{\rm c}$ (see Fig.~\ref{fig4}a). 

A second route is to change the nature of the connector. Indeed, a direct coupling $xJ$ of stars by vertical dimers also protects the UQP ground state for $x$ not too large. A graphical illustration for the case $N_{\rm t}=3$ is shown in Fig.~\ref{fig4}b. Again, excitations inside the star and the local triplon centered on the connecting dimer will be exact excitations of the full problem. 

As $x$ increases, a phase transition will certainly occur. Details depend on the nature of the connector. For the case of a direct star coupling just discussed this will likely be the triplon mode on the connector. If true, a magnetically ordered triplon condensate will be stabilized for large $x$. This triplon mode can be exactly captured by studying a cluster of two neighboring stars coupled by a single dimer. We find a critical value $x_{\rm c}\approx 0.75$ for $N_{\rm t}=3$.     

In contrast, for the connector comprised of the 1d segment of a Shastry-Sutherland chain discussed first, a different scenario is likely. One knows for the Shastry-Sutherland model that triplons strongly attract in the sector with $S$ $=$ $0$ when placed on neighboring vertical dimers, meaning the two ends of a connector in our case \cite{Knetter00b}. We find strong evidence for this scenario for coupled fully-frustrated plaquettes. Here, one observes on a cluster of two plaquettes and one connector that the UQP becomes unstable for $x_{\rm c}\approx 0.85$. Instead of a single triplon, it is again a singlet state keeping the ground state degeneracy of the UQP phase. The two triplons on neighboring cages bind into a singlet bound state.  

We stress that the above considerations do not fix the nature of the phase transition. It might well be that the transition for both cases turn out to be first order as it is known for the one-dimensional Shastry-Sutherland chain \cite{Ivanov97}. If so, the phase transition point would tend slightly to lower values of $x$.

%Figure Extensions
%%%%%%%%%%%%%%%%%%%%%%%%%%%%%%%%%%%%%%%%%%%%%%
%%%%%%%%%%%%%%%%%%%%%%%%%%%%%%%%%%%%%%%%%%%%%%

% 
%
%%%%%%%%%%%%%%
% Relation to Kagome GS
%%%%%%%%%%%%%%
%
%

Next, we want to discuss a connection of the case $N_{\rm t}=6$ with Shastry-Sutherland connector to the strongly debated Heisenberg model on the Kagome lattice. There are two possible close packings of chiral quantum cages on the Kagome topology (see Fig.~\ref{fig4}c). The first packing is such that honeycombs remain between the cages. {\it Assuming} a spontaneous dimerization, one recovers the valence bond solid consistent with 36-site unit cell proposed to be linked to the true ground state of the Heisenberg model on the Kagome lattice \cite{Zeng95,Singh07}. The second packing is denser but leaves dangling spins which is clearly unpreferable. But removing these spins yields a {\it depleted} Kagome lattice. Quite remarkably, one can show that the depleted model has VBS order with a 24-site unit cell. The easiest way to see this is by identifying the chiral pseuodspin states as {\it exact} eigenstates of a single cage. In 2nd order inter-cage degenerate perturbation theory, the effective pseudospin model turns out to be precisely an Ising ferromagnet in a transverse ``magnetic'' field (see also Ref.~\onlinecite{Syromyatnikov02}): 
$H_{eff}=-j\sum_{\mu,\nu}T_{\mu}^{z}T_{\nu}^{z} -h\sum_{\mu}T_{\mu}^{x}$, with $j,h<<J$ .
%\footnote{Details will be presented elsewhere.}.  
The ``field'' stabilizes a VBS state with ``ferro'' chiral order. The elementary excitations are therefore weakly dispersive ``pseudospin'' flips involving co-operative flipping of the six dimers on the internal star.  
%For $h<<j$, these read $E(k)=2j\left[cos k_x+cos k_y cos \left( (k_x + \sqrt{3}k_y)/2\right)\right}]-h$ in the semiclassical approximation. 
This succinctly illustrates the crucial influence of the unusual ground states found here on geometry, with disordered valence bond liquid, valence bond solid and magnetically ordered phases arising for differing geometries.

% 
%
%%%%%%%%%%%%%%
% Specific heat
%%%%%%%%%%%%%%
%
%

Let us finally discuss the specific heat. For $x$ $<$ $x_{\rm c}$, the decoupled (disordered) phase is characterized by a specific heat entirely due to the fluctuating chiral doublet, giving $C(T)\simeq T^{-2}$, reflecting the finite $T$=$0$ entropy per site.  In the case where magnetic order arises for $x$ $>$  $x_{\rm c}$, quenching of this entropy will yield $C(T)$ $\simeq$ $T^{2}$.  For the depleted Kagome magnet, two weakly dispersive bands of singlets, defined by $E_k^{\pm}=\pm(\sqrt{\epsilon_k^2+\Delta_k^2}-h)$ with $\epsilon_k=2j\left[\cos k_y+2\cos \left( \sqrt{3}k_x /2\right) \cos \left( k_y /2\right) \right]$ and $\Delta_k=(2/3)j\sin (k_y/2)\left[\cos (\sqrt{3}k_x/2)-\cos (k_y /2)\right]$ for $h<<j$, give rise to phonon-like excitations with velocity $v\simeq j$, and, in two dimensions, this will yield $C(T)\simeq \exp(-h/t)$ for $T\rightarrow 0$ and $C(T)\simeq T^{2}$ for higher temperatures.

% 
%
%%%%%%%%%%%%%%
% Conclusion and Discussion
%%%%%%%%%%%%%%
%
%

In conclusion, we have presented families of frustrated quantum magnets with hitherto undiscovered gapped SL ground states. The ground state as well as elementary excitations can be determined exactly and both possess a macroscopic degeneracy at $T=0$. One important question is the fate of the finite entropy upon heating these systems. Indeed, thermal fluctuations induce a finite triplon density on each cage. These triplons correspond to different effective pseudo-spins (not to be confused with the chiral pseudo-spins $T_\mu$). This leads to an inter-cage coupling, producing an ordered phase continuously connected to that found at $T$ $=$ $0$ for $x$ $>$ $x_{\rm c}$. 

For experimental realizations, we only demand flexibility on the lattice design since our models solely contain nearest-neighbor Heisenberg couplings. We suggest that experimental realizations of QCA in suitable geometries, artificially fabricated JJA, or trapped ions may exhibit some of the unusual features found in this work.  Arrays of magnetic clusters, e.g, of Cu$_{3}$ triangles suitably engineered on a substrate, may also reveal parts of the exotica proposed here. In particular, given the protected chirality, an exciting option for future study is to use this aspect in a setting favorable for quantum computation in ``interaction free subspaces'' \cite{Zhou02} to minimize the troublesome decoherence problem in quantum information.
 
\acknowledgments KPS acknowledges ESF and EuroHorcs for funding through his EURYI. We thank F.~Mila, J.~Stolze, and G.~S.~Uhrig for fruitful discussions.

\end{document}